\documentclass[%
 reprint,
 amsmath,amssymb,
 aps,nofootinbib,   
]{revtex4-1}

\usepackage{bm}% bold math
\PassOptionsToPackage{linktocpage}{hyperref}
\usepackage[hyperindex,breaklinks]{hyperref}
\usepackage{enumitem}
\usepackage{slashed}

\usepackage{xcolor}
\usepackage{float}

\usepackage{array}
\usepackage{mathtools}

\usepackage{etoolbox}

\begin{document} 

\title{The information of the information paradox: On the quantum information meaning of Page's curve.}
 
\author{  C\'esar G\'omez} 
\affiliation{Instituto de F\'{i}sica Te\'orica UAM-CSIC, Universidad Aut\'onoma de Madrid, Cantoblanco, 28049 Madrid, Spain}

%\date{\today}

\begin{abstract}
Page's curve for the fine grained entropy of the black hole radiation is obtained once we include the effect of a non vanishing quantum Fisher information in the evaporation data. This quantum Fisher information scales as the inverse of the original black hole entropy. The modified fine grained entropy can be naturally related to a relative entanglement entropy with the quantum Fisher information defining the curvature at Page's point.

 \end{abstract}
\maketitle

%\pacs{(old) 14.60.Pq,13.15.+g,04.60.-m,11.30.Rd}

%\section{Introduction}
After the seminal paper of Page \cite{Page} Hawking's information paradox \cite{Hawking} is presented as the mismatch between the monotonous growth of the entropy of radiation predicted by Hawking and the expected Page curve for the radiation entropy if the process is governed by unitary evolution. For a black hole of Bekestein Hawking entropy $N$ Hawking's growth of the entropy of radiation fits qualitatively with the Page curve for times $t<t_P$ where Page time corresponds to the time at which the original black hole has emitted around $N/2$ quanta i.e. the time at which the Bekestein Hawking entropy is equal to the radiation entropy. After this time Page's curve starts to decrease in full contrast with Hawking's expectations. In more precise terms, Page's time is the time at which the fine grained entropy of the radiation reaches its maximum. The paradox makes explicit the problem to identify what can goes wrong in Hawking's original semiclassical computation. This is paradoxical since, in principle, the semiclassical approximation should be reliable at this time where the black hole can be as macroscopic as wished.

In this note I will offer a quantum information approach to the information paradox and a natural way to get Page's curve once quantum information, in particular quantum Fisher information, in taken into account. This approach is framed in the context of a series of recent papers \cite{Gomez} on physics applications of quantum estimation theory \cite{Paris}. The main result of this note is summarized in equation (\ref{formula}).

A key ingredient in Page's construction is to model the evaporation of the black hole working with a total Hilbert space of finite dimension of order $2^N$. The evolution leads to a time dependent decomposition of this Hilbert space into two pieces; the remaining black hole at each time and the corresponding emitted radiation. Tracing over the black hole interior leads to a density matrix $\rho_{rad}(t)$ for the radiation and to the entanglement entropy $S_{rad}(t) = - tr (\rho_{rad}(t) \ln \rho_{rad}(t))$. This is the entropy we expect to follow Page's curve in time. Hawking's answer for $S_{rad}(t)$ is in first approximation defined as the entanglement entropy of the set of entangled pairs created at time $t$ and therefore grows with $t$. Let us denote this entropy $S_{Hawking}(t)$. 

So what is the wrong assumption in Hawking's computation? The simplest answer is that the model $\rho_{Hawking}(t)$ of the state $\rho_{rad}(t)$ defined by tracing over the inner partners of a set of independent entangled pairs is simply too crude to reflect the real situation. In this representation of $\rho_{rad}(t)$ we are ignoring all the features of the black hole interior with the exception of the set of Hawking partners created after a time $t$ of evaporation. To figure out how these missing effects of the black hole interior can reproduce Page's curve is still an open question. 

Recently it has been announced important progress in reproducing Page's curve \cite{Page2, Page3, Page4} on the basis of a modified definition of the fine grained entropy of the radiation motivated by generalizations of Ryu Takayanagi (RT) \cite{RT,RT2,RT3,RT4,RT5,RT6,RT7,RT8} formula. This modification of the entropy requires to define associated with each time $t$, as measured by an external observer, an spacelike surface $\Sigma(t;r_i(t))$ depending on a set of positions in the black hole interior {\it island coordinates}. For the case of just one island this one is defined at each time $t$ by the hyper surface going from $r=0$ to $r(t)$.  The correct radiation entropy is now defined by the extremal of a semiclassical functional $S(\Sigma(t;r(t))$ inspired by RT formula and justified by the euclidean path integral and the replica trick. The so defined entropy contains three pieces: i) the entropy of radiation associated with the island, ii) the entropy of radiation associated with the hypersurface going from the location of the observer at time $t$ and the spacial infinity and iii) the analog of the thermodynamic entropy for the two dimensional surface of radius $r(t)$. The semiclassical extremal of this functional is shown to reproduce Page's curve. In essence the role of the island in the black hole interior is to account for a partial purification of the emitted radiation.

A similar discussion can be developed for the black hole entropy during evaporation. In this case Page's curve differs from the Hawking Bekestein curve describing the time evolution of the black hole thermodynamic entropy for $t<t_P$. The modified entropy is defined using a hypersuface going from the location of the observer at time $t$ to a point in the black hole interior with radial coordinate $r(t)$. The modified entropy contains two pieces one representing the semiclassical radiation entropy on this hypersurface and another piece representing the thermodynamic entropy of the two dimensional surface of radius $r(t)$ \footnote{See \cite{Malda} and references therein for an excellent recent review of this construction.}. The very non trivial ingredient of this construction is to find  a semiclassical formula for the entropy, based on saddle point dominance for the gravity partition function, that fits Page's unitarity condition. However the main problem of the construction is that it does not provides direct information about the quantum density matrix of the radiation whose fine grained von Neumann entropy follows Page's curve.

The new approach to the information paradox based on the modified RT entropy formula is framed in \cite{Malda} in what is presented as {\it the central dogma}. In essence the so called {\it central dogma} identifies the black hole, as view from outside, as an ordinary Hamiltonian quantum system with a {\it finite dimensional Hilbert space} \footnote{Semiclassically this finiteness condition simply means that the classical phase space is compact with finite volume in $\hbar$ cell units. It is customary to describe quantum systems with finite dimensional Hilbert space in terms of a finite set -- given by the logarithm of the dimension -- of q-bits and to refer to these q-bits as effective degrees of freedom. }. This central dogma is very much in agreement with the black hole portrait suggested in \cite{Gia} where the black hole is modeled as a self sustained quantum Bose condensate of gravitons at criticality with the description from outside being determined by the dynamical depletion properties of the condensate. The results presented in this note on the entropy of radiation although independent on the details of the portrait lead, when combined with the main portrait depletion equation, to the explicit time dependence of the radiation entropy.

Something important, but many times ignored, for the correct understanding of Page's curve is the associated curve of information. At each time we can define, following \cite{Page}, a qualitative measure of the information contained in the radiation as
\begin{equation}
I(t) = \ln (d_{rad}(t)) - S_{rad}(t)
\end{equation}
where $d_{rad}(t)$ is the dimension of the Hilbert space of the radiation at time $t$. We observe already from this simple definition of information that, if we follow Hawking's growth, the information remains constant and vanishing during the whole evaporation process. By contrary if we consider Page's curve the information starts to increase after Page's time until reaching its maximal value at the end of the evaporation. Hence it looks clear that the main problem with Hawking's description of the radiation is that is not accounting for any form of growth of the information contained in the radiation.

For the density matrix $\rho_{Hawking}(t)$ defined by a set of $N_e(t)$ maximally entangled and independent pairs, the entanglement entropy obtained after tracing over the inner partners is simply given by
\begin{equation}
S_{Hawking}(t) = N_e(t)
\end{equation}
where $N_e(t)$ represents the number of emitted Hawking quanta at time $t$ measured taking as $t=0$ the moment of black hole formation. 

{\it The main claim of this note is that the fine grained entropy of radiation is given by the following general formula
\begin{equation}\label{main}
S_{rad}(t) =  S_{Hawking}(t) - I_{F}(t)
\end{equation}
with $I_{F}(t)$  fully determined by the quantum Fisher information of the radiation} \footnote{The thermodynamical meaning of the former formula can be understood assuming that whenever the black hole starts to deliver information the emitted radiation can be used to do some thermodynamic work and that the correct entropy should account for the corresponding free energy contribution.} .

In order to define $I_{F}(t)$ we shall use quantum estimation theory. Let us identify as external clock the number $N_e(t)$ of emitted quanta. For the correct density matrix of the radiation $\rho_{rad}(t)$ the number of emitted Hawking quanta $N_e(t)$ works as an external time parameter i.e. $\rho_{rad}(N_e(t))$. Let us define the corresponding estimator operator $\hat N_e$ such that $tr(\rho_{rad}(t) \hat N_e) \equiv \langle \hat N_e \rangle = N_e(t)$. The corresponding quantum Fisher information is defined by
\begin{equation}\label{Fisher}
F^{-1}= \langle \hat N_e ^2 \rangle - \langle\hat N_e\rangle \langle \hat N_e \rangle
\end{equation}
Now we will define $I_{F}(t)$ as
\begin{equation}
I_{F}(t) = F N_e(t)^2
\end{equation}
Using now the main formula (\ref{main}) we get our basic result, namely
\begin{equation}
S_{rad}(t) = N_e(t) - FN_e(t)^2
\end{equation}
It remains to set the value of $F$. If the time evolution is unitary we expect $F$ to be time independent. The definition of $F$ associated with $N_e$ is given by
\begin{equation}
F = Tr (\rho_{rad}(N_e) \hat L_{Ne}^2)
\end{equation}
with $\hat L_{N_e}$ the operator defined as usual by
\begin{equation}
\frac{d\rho_{rad}(N_e)}{dN_e} = \frac {\hat L_{N_e} \rho_{rad}(N_e) + \rho_{rad}(N_e) \hat L_{N_e}}{2}
\end{equation}
Thus $F^{-1}$
is essentially the standard deviation of the estimator $\hat N_e$ as given in (\ref{Fisher}). To define this standard deviation we can think in an ensamble of identical black holes and to measure the observable {\it estimator} $\hat N_e$. However and irrespectively of the purely statistical deviation, the quantum Fisher function gives us the {\it intrinsic quantum standard deviation} for the quantum estimator. We will {\it conjecture} as the natural value for $F$
\begin{equation}
F= \frac{1}{N}
\end{equation}
with $N$ the Bekestein Hawking entropy of the original black hole. The heuristic reading of this identification is based on the natural interpretation of $F$ for the original black hole as the typical quantum energy uncertainty $\Delta^2(E) \sim 1/N$.

Using these ingredients we get as our model for the exact radiation entropy:
\begin{equation}\label{formula}
S_{rad}(t) = N_e(t) - \frac{1}{N} N_e(t)^2
\end{equation}
The previous construction provides a general recipe to define the fine grained entropy. For instance for the black hole itself an external observer will use as estimator the thermodynamic black hole entropy $N_{BH}(t)$. The exact black hole entropy will be defined for a density matrix $\rho_{BH}(N_{BH})$ whose dependence on $N_{BH}$ is again set by the quantum Fisher function $F$. The former recipe for defining this entropy leads to
\begin{equation}
S_{BH}(t) = N_{BH}(t) - F N_{BH}(t)^2 = N_{BH}(t) - \frac{1}{N} N_{BH}(t)^2
\end{equation}
that follows Page's curve as expected. The purely quantum contribution to this entropy is the information component  $F N_{BH}(t)^2$. Defining the quantum estimator $\hat N_{BH}$ we get as before $F^{-1} = \langle \hat N_{BH} ^2 \rangle - \langle\hat N_{BH}\rangle \langle \hat N_{BH} \rangle$.

The skeptic reader can focus on (\ref{formula}) and ignore our quantum information justification. 
Let us now briefly discuss some key aspects of this entropy formula (\ref{formula}):
\begin{itemize}
\item It reproduces Page's curve. Indeed for times for which $N_e(t) << N$ we get $S_{rad}(t) \sim N_e(t)$ in agreement with Hawking. Moreover for $N_e \sim N$ the radiation entropy $S_{rad}(t)$ decreases as the black hole thermodynamic entropy. The curve, in $N_e$ coordinates, has a maximum at  $N_e=\frac{N}{2}$. 
\item To define the explicit time dependence of $N_e(t)$ we can use the main formula describing depletion in the black hole portrait \cite{Gia1}, namely
\begin{equation}
\frac{dN_e(t)}{dt} = \frac{1}{\sqrt{N_e(t)}}
\end{equation}
leading to
\begin{equation}
S_{rad}(t) = t^{2/3} - \frac{1}{N} t^{4/3}
\end{equation}
with $t_P$ in Planck units given by
\begin{equation}
t_P = (\frac{N}{2})^{3/2}
\end{equation}
\item The information $I_{F}(t)$ goes as
\begin{equation}
I_{F}(t) = \frac{1}{N} t^{4/3}
\end{equation}
This information becomes maximal and equal $N$ after complete evaporation i.e. after a time $O(N^{3/2})$ in Planck units.

\item Comparing with the island formula for $r(t)$ the location of the island in the black hole interior, we get, at extremality
\begin{equation}\label{eW}
\frac{\pi r^2(t)}{G_N} + S_{island}(t) = - \frac{N_e(t)^2}{N}
\end{equation}
which makes explicit the purifying role of the island contribution.
\item A related natural question that we will leave unanswered  is the connection between RT like formulas and the former description. Probably the secret lies in the connection of information metric and gravity dual metric 
\cite{Taka} ( see also \cite{johana} ). Concerning this important issue, namely the connection between the semiclassial entropy of the entanglement wedge of the radiation and formula (\ref{eW}), we have nothing substantial to say in this note.
\end{itemize}

It could be instructive, in order to get a more clear physical picture of Page's time, to define formally the function $\tilde S(N_e) \equiv -S_{rad}(N_e)$ i.e.
\begin{equation}
\tilde S(N_e) = FN_e^2 -N_e
\end{equation}
At Page's time we get the analog of the first entanglement law for $\tilde S$, namely $\frac{d \tilde S}{dN_e} =0$. Moreover the quantum Fisher information becomes simply $F= \frac{d^2\tilde S}{dN_e^2}$. These are the equations we should expect for $\tilde S$ defining the relative entanglement entropy with the Page point describing a sort of equilibrium \footnote{With the two components of the finite dimensional Hilbert space having the same dimension.}. In other words the fine grained entropy of the radiation can be defined as minus the relative entanglement entropy relative to Page's point.

The main lesson we can extract from the previous discussion is that the correct entropy of the radiation accounts for the information through a quadratic dependence in the number of Hawking emitted quanta with coefficient given by the quantum Fisher information. This quantity is introduced in (\ref{Fisher}) and it is identified with $1/N$ for $N$ the black hole entropy. The key point is that, as advertised several times in the black hole portrait,  Hawking's computation is effectively done in the $N=\infty$ limit or equivalently $F=0$. However is also very important to notice that the quantum effect defined in (\ref{Fisher}) is not semiclassical -- which will implies an exponentially suppressed quantum Fisher information -- but purely quantum \footnote{Superficially it could be tempting to interpret the non vanishing quantum Fisher information as the effect of some gravitational correlation induced by semiclassical wormholes. This will lead to an exponentially suppressed quantum Fisher information that in our frame will be unable to reproduce Page's curve.}. In essence and from a simple minded thermodynamic point of view what goes wrong in Hawking's computation is the assumption that the black hole evaporation products cannot be used to do any {\it work}. What is interesting is the observation that unitary Hamiltonian evaporation is only possible if the evaporation products can be used to do some net work, a work that shows up in the form of non vanishing quantum information \footnote{The origin of the long standing confusion lies in the fact that Hawking's analysis is actually done, as originally pointed out in \cite{Gia1}, in a very peculiar double scaling limit $M=\infty$ and $L_P=0$ with $ML_P^2$ finite and with non vanishing $\hbar$. In this limit we have Hawking radiation with vanishing work i.e. $F=0$.}.

\acknowledgments
 This work was supported by grants SEV-2016-0597, FPA2015-65480-P and PGC2018-095976-B-C21.

\end{document}